\documentstyle[aps]{revtex}
\draft
\begin{document}

\twocolumn[\hsize\textwidth\columnwidth\hsize\csname
@twocolumnfalse\endcsname

\title{Gravitational Collapse of Dust with a Cosmological Constant
}
\author{Kayll Lake}
\address{ Department of Physics, Queen's University, Kingston, Ontario, Canada, K7L 3N6}

\date{\today}
\maketitle
\begin{abstract}
The recent analysis of Markovic and Shapiro on the effect of a cosmological constant
on the evolution of a spherically symmetric homogeneous dust ball is extended to include
the inhomogeneous and degenerate cases. The histories are shown by way of effective potential
and Penrose-Carter diagrams. 
\end{abstract}

\vskip2pc]

% \narrowtext

% \twocolumn

\section{Introduction}

Recently, Markovic and Shapiro \cite{MS}, motivated by some observational suggestions of a positive cosmological constant \cite{Lambdaobs}, have reexamined the effect of this constant on the evolution of a homogeneous dust ball embedded in vacuum. This paper extends their analysis so as to include the inhomogeneous and degenerate cases. The qualitative behavior of the boundary histories are shown by way of effective potential and Penrose-Carter diagrams. The case $\Lambda < 0$ is included as it provides for an interesting contrast. The well known case $\Lambda = 0$ is not included.

\section{Dust}

The study of spherically symmetric distributions of matter without pressure in the general theory of relativity has a long history. It is fair to say that the dynamics of this ``Lema\^{i}tre - Tolman - Bondi" solution are well understood, even with a non-vanishing cosmological constant \cite{omer}. Whereas the discovery of ``shell-focusing" singularities in dust added a new dimension to the dynamics \cite{ES}, these singularities are now well studied \cite{lake1} and are not considered here.

We review the dynamics to set the notation. First, recall that the flow lines of all dust distributions are \textit{geodesic}. As a consequence, with spherical symmetry we can choose synchronous comoving coordinates $(\textsf{r},\theta,\phi,\tau)$ so that the line element associated with the dust takes the form
\begin{equation}
\label{dust}
ds^2 = e^{\alpha(\textsf{r},\tau)}{d\textsf{r}^2} + R(\textsf{r},\tau)^2(d\theta^2 + \sin^2 \theta  d\phi^2)-d\tau^2.
\end{equation}
As long as $R^{'} \not= 0 ~(^{'}\equiv \frac{\partial}{\partial \textsf{r}})$ \cite{ruban} we obtain
\begin{equation}
\label{alpha}
e^{\alpha(\textsf{r},\tau)} = \frac{{R^{'}}^{2}}{1+2 E(\textsf{r})}.
\end{equation}
A further integration gives one more independent function of $\textsf{r}$
\begin{equation}
\label{M}
{{R^{*}}}^{2} - 2 E(\textsf{r})-\frac{\Lambda R^2}{3} = \frac{2 M(\textsf{r})}{R}
\end{equation}
where $^{*}\equiv \frac{\partial}{\partial \tau}$. The energy density follows as
\begin{equation}
\label{rho}
4 \pi \rho(\textsf{r},\tau)  = \frac{M^{'}}{R^{2}R^{'}}.
\end{equation}
Many explicit forms of $R(\textsf{r},\tau)$ are known, but these are not of interest here.

\section{Vacuum}

The $\Lambda$ generalization of the Schwarzschild vacuum is well known. In terms of familiar
curvature coordinates $(r,\theta,\phi,t)$ the line element is given by
\begin{equation}
\label{vacuum}
ds^2 = \frac{dr^2}{f(r)} + r^2(d\theta^2 + \sin^2 \theta  d\phi^2)-f(r)dt^2,
\end{equation}
where
\begin{equation}
\label{f}
f(r) = 1-\frac{2 m}{r}-\frac{\Lambda r^2}{3}.
\end{equation}
The associated generalization of the Birkhoff theorem is well known \cite{bonnor}. It is interesting to note that the $\Lambda$ generalization of the Israel theorem \cite{israel} is not known. Geodesically complete forms of the metric (\ref{vacuum}) along with Penrose - Carter diagrams are now well know \cite{lake2}.

The coordinates $(r,\theta,\phi,t)$ are adapted to two Killing vectors and so geodesics of the metric (\ref{vacuum}) have two constants of motion. The orbits are stably planar and we choose the plane to be $\theta = \pi/2$. The momentum conjugate to $\phi$ is the orbital angular momentum  \textit{l},
\begin{equation}
\label{l}
r^2\dot{\phi} = \textit{l},
\end{equation}
and the momentum conjugate to $t$ is the energy $\gamma$,
\begin{equation}
\label{energy}
f(r)\dot{t} = \gamma.
\end{equation}

For timelike geodesics we can take $^. = \frac{d}{d\lambda}$ where $\lambda$ is the proper time. In what follows we are interested in radial motion so that $\textit{l} = 0$. $\gamma$, however, plays a central role. The timelike geodesic equations reduce to
\begin{equation}
\label{geodesics}
\gamma^2 - \dot{r}^2 = -f(r).
\end{equation}
We can write $P(r) \equiv -f(r)$ and treat $P$ as the effective potential of elementary mechanics.

\section{Junction}

The junction of dust and vacuum in spherical symmetry by way of the Darmois - Israel conditions is well understood \cite{ML}. To summarize, the continuity of the first fundamental form associated with the boundary ($\Sigma$) ensures that the continuity of $\theta$ and $\phi$ in metrics (\ref{dust}) and (\ref{vacuum}) is allowed and that the history of the boundary is given by
\begin{equation}
\label{history}
R(\textsf{r}_{\Sigma},t) = r_{\Sigma}.
\end{equation}
The continuity of the second fundamental form guarantees that the flow lines of the boundary particles are simultaneously geodesic of both enveloping 4-geometries. The junction conditions demand that 
\begin{equation}
\label{msurf}
M(\textsf{r}_{\Sigma}) = m,
\end{equation}
and that for $R^{'} \not= 0$
\begin{equation}
\label{gammab}
E(\textsf{r}_{\Sigma}) = \frac{\gamma^2-1}{2}.
\end{equation}
The case $R^{'} = 0$ gives $\gamma = 0$.

\section{Discussion}

The qualitative history of the geodesics of (\ref{vacuum}), and via (\ref{gammab}) therefore of the dust boundary $\Sigma$, can be obtained from a sketch of $P$ (and in particular the requirement that $\gamma^2 \geq P$). These are shown in Figure 1. (The roots $(r_0,r_2,r_3)$ are given explicitly in \cite{lake2}.)
Note that for $\Lambda < 0$ \textit{all} orbits are closed, in contrast to $\Lambda \geq 0$. The case $\gamma = 0$ is unique in the sense that $\Sigma$ traverses the bifurcation of the Killing horizons (in the non-degenerate cases).

The Penrose - Carter diagrams are shown in Figure 2. The possible histories of $\Sigma$ are shown. The dust can be matched to the left or to the right. The degenerate case $3 m = 1/\sqrt{\Lambda}$ requires a special coordinate construction \cite{lake3}. Note that here the case $\gamma = 0$ is associated with unstable equilibrium at the points of internal infinity.

\begin{figure}
\caption{
Sketches of the effective potential $P$ for timelike geodesics. Choose $\gamma^2 \geq 0$ and note that $\gamma^2 \geq P$. $\overline{r}\ \equiv (\frac{3 m}{\Lambda})^{1/3}$. The convention for the Lagrangian is $2L = -1$.
}
\end{figure}

\begin{figure}
\caption{Penrose - Carter diagrams for the trajectories shown in Figure 1. In all cases the choice of future can be reversed.}
\end{figure}

\subsection*{Acknowledgments}
This work was supported by a grant from the Natural Sciences and Engineering Research Council of Canada. I would like to thank Sean Hayward for pointing out the work by Nakao and Jos\'{e} Lemos for reminding me of his work on Oppenheimer-Snyder collapse.


\begin{references}
\bibitem{ruban} Following Ruban (V. A. Ruban, JETP {\bf 56}, 1914 (1969)), the case $R^{'} = 0$ can be examined as a limit of the subclass $R^{'} = \Gamma(\textsf{r})e^{\alpha/2}$. In this case the junction conditions impose the restriction $\Gamma(\textsf{r}_{\Sigma}) = \gamma$.
\end{references}
\end{document}